\documentclass[reprint,amsmath,amssymb,aps]{revtex4-1}

\usepackage{epstopdf}
\usepackage{graphicx}
\usepackage{siunitx}
\usepackage{amsmath}
\usepackage{makecell}
\usepackage{epsfig}

\begin{document}
\title{Fine-structure changing collisions in $^{87}$Rb upon D2 excitation in the hyperfine Paschen-Back regime}

\author{Clare R Higgins, Danielle Pizzey and Ifan G Hughes}

\affiliation{Department of Physics, Durham University, South Road, Durham, DH1 3LE, UK}

\affiliation{Corresponding author: danielle.boddy@durham.ac.uk}

\begin{abstract}
We investigate fine structure changing collisions in $^{87}$Rb vapour upon D2 excitation in a thermal vapour at 350~K; the atoms are placed in a 0.6~T axial magnetic field in order to gain access to the hyperfine Pashen-Back regime. Following  optical excitation on the D2 line, the exothermic  transfer 5P$_{3/2}$~$\rightarrow$~5P$_{1/2}$ occurs as a consequence of buffer-gas collisions; the $^{87}$Rb subsequently emits  a photon on the D1 transition. We employ single-photon counting apparatus to monitor the D1 fluorescence, with an etalon filter to provide high spectral resolution.  By studying the D1 fluorescence when the D2 excitation laser is scanned, we see that during the collisional transfer process the $m_{J}$ quantum number of the atom  changes, but the nuclear spin projection quantum number, $m_{I}$, is conserved. A simple kinematic model incorporating a coefficient of restitution in the collision accounted for the change in velocity distribution of atoms undergoing collisions, and the resulting fluorescence lineshape.
The experiment is conducted with a nominally ``buffer-gas free" vapour cell; our results show that fine structure changing collisions are important with such media, and point out possible implications for quantum-optics experiments in thermal vapours producing entangled photon pairs with the  double ladder configuration,   and solar physics magneto-optical filters.

\end{abstract}

%
%
%
\maketitle
%
%

\section{Introduction}
 Hot atomic vapours are work horses of modern atomic physics experiments. These media are ideally suited  for quantum-optics and atom-light interaction experiments as they combine: a large resonant optical depth; long coherence times; and well-understood atom–atom interactions~\cite{pizzey2022laser}.
Thermal atomic vapours find utility in applications spanning from the fundamental: electromagnetically induced transparency~\cite{finkelstein2023practical, zhang2024interplay}; nonlinear and quantum optics~\cite{glorieux2023hot}; cooperative effects in confined geometries~\cite{alaeian2024manipulating}; through to the applied: magnetometry~\cite{fabricant2023build}; THz imaging~\cite{downes2023practical}; narrowband optical filters~\cite{uhland2023build}.  A noteworthy feature of using thermal atomic vapours is the simplicity of both the experimental set up and modelling  few-level atom-light interactions~\cite{downes2023simple}. One branch of quantum optics where thermal vapours find great utility is that of producing heralded entangled photon pairs~\cite{willis2010correlated, willis2011photon, shu2016subnatural, davidson2023bright, kim2024collective}. 

The spectroscopic investigation of the interactions of alkali-metal vapours with buffer-gases (such as inert gases, ${\rm N}_2$) has been thoroughly studied historically; see, for example, the textbooks~\cite{corney1978atomic, thorne1988spectrophysics} for an overview. The collisional relaxation of excited state population and the concomitant optical line broadening and shift are understood comprehensively experimentally and theoretically~\cite{lewis1980collisional}.
In many coherent population trapping and magnetometry experiments--including chip-scale atomic clocks--in miniaturised vapour cells a limiting feature can be the ground state coherence relaxation time being limited by collisions of the alkali-metal atoms with the cell wall. The standard approach to reduce the relaxation rate is to add buffer-gas to the cell, as the diffusion rate of the atoms out of the laser beam towards the cell walls is reduced~\cite{brandt1997buffer, schwindt2004chip, knappe2004microfabricated, vanier2005atomic, wang2014review}. There is a trade off, with higher buffer-gas pressure leading to too large a collisional rate, leading to an optimal gas pressure for optimising the contrast of narrow resonances~\cite{sargsyan2024influence}. There is much current interest in realising vapour cell devices with functionalised cells~\cite{raghavan2024functionalized} and micro-machined deep silicon atomic vapour cells~\cite{dyer2022micro}. Recent work has demonstrated  nitrogen buffer-gas pressure tuning in a micro-machined vapour cell~\cite{dyer2023nitrogen}. Another example of enhancing the application by adding a buffer-gas is when using cascaded atomic magneto optical filters~\cite{logue2022better} to monitor the solar magnetic field~\cite{ refId0, refId1, refId2} where optimising line shapes by shifting and broadening lines is routine.  A field that exploits the fine-structure changing collisions of alkali-metal atoms when subject to high buffer-gas collisions is that of generating high-powered lasers with diode-pumped alkali lasers~\cite{krupke2012diode, gao2013review, pitz2017recent}.

There are examples where the collisions associated with a buffer gas can have deleterious effects on the desired performance; for example, the degradation of the figure-of-merit in Faraday filters~\cite{zentile2015optimization}. 
The nonlinear process of nondegenerate four wave mixing (FWM) can be employed, with two driving fields producing an entangled photon pair.
Such a system has been widely studied in different geometries: the ``double lambda''~\cite{boyer2007ultraslow, boyer2008entangled, camacho2009four, kim2018generation}, ``double ladder''~\cite{davidson2021bright, hsu2008controlled, khadka2012four, lee2016highly} and ``diamond'' schemes~\cite{willis2009four, walker2012trans, wen2014electromagnetically, offer2018spiral, offer2021gouy}.  In Rb a well-studied double ladder scheme involves 5S$_{1/2}-$5P$_{3/2}-$5D$_{5/2}$ or 5S$_{1/2}-$5P$_{1/2}-$5D$_{3/2}$ transitions, whereas the diamond scheme uses 5S$_{1/2}-$5P$_{3/2}-$5D$_{1/2}$~\cite{Whiting:FWM, mathew2021single, higgins2023quantum}. State-changing collisions can  degrade the performance of thermal vapour single-photon sources~\cite{mathew2021single, higgins2023quantum}. 

The study of fine-structure changing collisions in a thermal vapour with excitation of $^{87}{\rm Rb}$ atoms  with  D2 (780~nm) light and monitoring D1 (795~nm) fluorescence performed in the hyperfine Paschen Back (HPB) regime is the topic of study of the present investigation.

In the  HPB regime the Zeeman shift exceeds the ground state hyperfine interaction. Recent work utilising the HPB regime span the fundamental~\cite{staerkind2023precision, mottola2023electromagnetically, mottola2023electromagnetically}, through to the applied ~\cite{mottola2023optical, staerkind2024high}. An estimate of the field needed to gain access to the HPB regime is $B_{\rm HPB}=A_{\rm hf}/\mu_{\rm B}$, where $A_{\rm hf}$ is the magnetic dipole constant for the ground term, and $\mu_{\rm B}$ is the Bohr magneton; this is evaluated to be 0.24~T for $^{87}$Rb~\cite{pizzey2022laser}. Numerous experimental studies have been performed in the HPB regime with Rb~\cite{tremblay1990absorption, Weller2012a, sargsyan2012hyperfine, Zentile2014a, sargsyan2014hyperfine,     Voigt5, briscoe2023voigt}. The spectroscopy of Rb is considerably easier to interpret when in the HPB regime because the separation of the optical transitions arising from the Zeeman interaction  exceeds the Doppler width, leading to  isolated atomic lines being  observed~\cite{higgins2021electromagnetically, WhitingEIA,  whiting2017single}. 
Of particular relevance to this study, investigating fine-structure changing collisions in the HPB regime makes the interpretation of the data significantly easier.

A schematic of our investigation is illustrated in Fig.\,\ref{fig:Fig1}~a). Upon optical excitation, buffer gas collisions can transfer the Rb atom from the 5P$_{3/2}$ state to the 5P$_{1/2}$, but also in the reverse direction, from 5P$_{1/2}$ state to the 5P$_{3/2}$ state. The first of these transfers, 5P$_{3/2}$~$\rightarrow$~5P$_{1/2}$, is an exothermic process, meaning that energy is transferred from the internal state of the atom to the kinetic energy of the colliding atoms. The second, 5P$_{1/2}$~$\rightarrow$~5P$_{3/2}$, is an endothermic process, meaning kinetic energy from the colliding atoms is transferred to the internal energy of the Rb atom, and subsequently the light. The second process is therefore energetically unfavourable, and happens at a lower rate.  
 
Previous experimental studies have comprehensively measured the cross-sections of these state changing collisions for a range of atomic species with a wide range of molecular collision partners~\cite{rotondaro1997collisional, rotondaro1998collision, rotondaro1998role}. Here, we do not attempt to reproduce these investigations; rather, we use a narrowband etalon filter~\cite{palittapongarnpim2012note, HigginsEtalon} to spectrally resolve the fluorescence.  In combination with non-degeneracy of energy levels provided by the large magnetic field, we gain further insight into fine-structure changing collisions. 

The rest of this paper is organised as follows: Section~\ref{sec:exp} outlines the experimental details; in Section~\ref{sec:results} the experimental results are presented; in Section~\ref{sec:model} we describe a simple kinetic collisions model that is used to predict the velocity distribution of the atoms after fine-structure changing collisions; in Section~\ref{sec:different} we present and analyse results obtained with a different buffer gas. Finally, we present our conclusions in Section~\ref{sec:conc}.

\section{Experimental Details}
\label{sec:exp}

\begin{figure}[htbp]
\centering
\includegraphics[width = \linewidth]{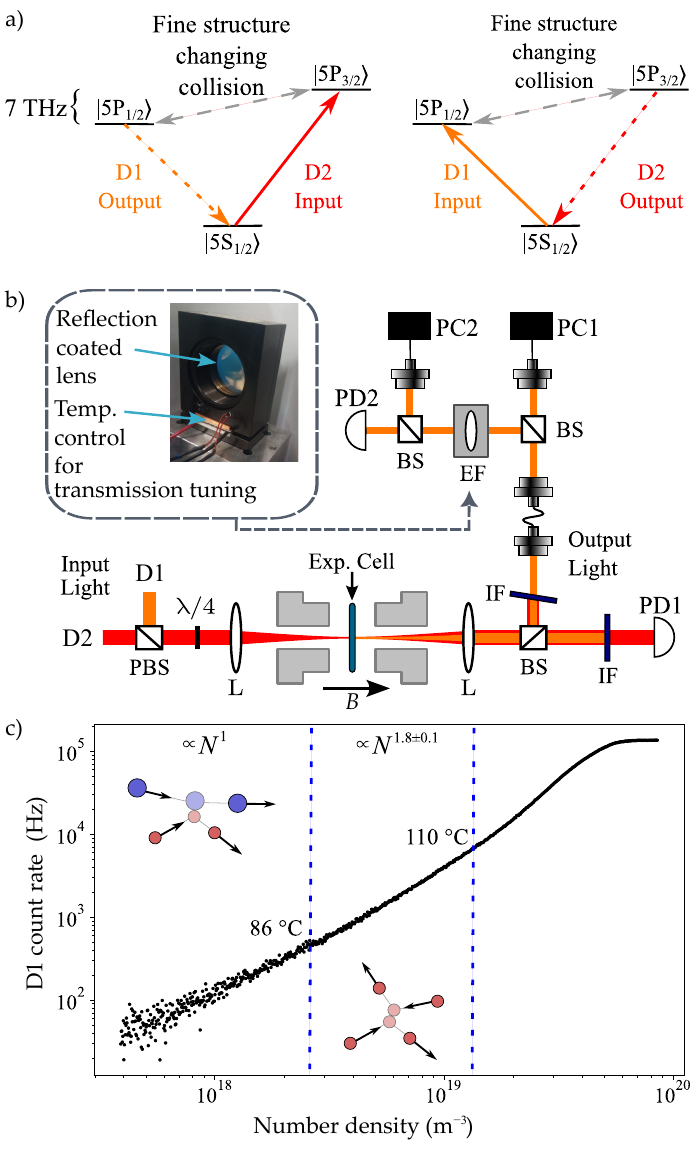}
\caption[Experimental setup for collisional transfer investigation.]{a) Rubidium energy levels. When D1 light is incident, D2 light is produced and vice versa; this is due to collisions transferring population between the two P states shown. b) Schematic of the experimental apparatus. The vapour cell is situated between magnets (grey blocks) producing a 0.6~T axial field. The input light can be either D2 (\SI{780}{\nm}) or D1 (\SI{795}{\nm}) laser light, whereas the output light consists of both D1 and D2 light. The output light is detected on two photon counters; one with and one without an etalon filter, necessary for high-resolution spectral filtering \cite{HigginsEtalon}. A photograph of the etalon filter is shown left of the experimental apparatus. PBS: Polarising beam splitter; $\lambda$/4: Quarter waveplate; L: Lens; Exp. Cell: \SI{2}{\mm} long 98\% $^{87}$Rb vapour cell; PD: Photodiode; BS: Beam splitter; IF: Interference filter; EF: Etalon filter. c) Number density dependence of \SI{795}{\nm} fluorescence rate when \SI{780}{\nm} light is incident on the vapour. At low number densities (and therefore temperatures), up to around \SI{86}{\celsius}, the D1 (\SI{795}{\nm}) fluorescence count rate increases linearly with number density; the dominating collisions in this regime are between Rb and buffer gas atoms, as pictorially represented by the colliding spheres, red and blue, respectively, inset. In the next region, between \SI{86}{\celsius} and \SI{110}{\celsius}, that relationship is close to quadratic ($\propto \text{$N$}^{1.8\pm0.1}$) due to Rb--Rb collisions (only red colliding spheres shown inset). Above \SI{110}{\celsius} the increase begins to level off, and then decrease, as the medium becomes optically thick, and fewer of the produced photons escape the medium and are detected. }
\label{fig:Fig1}
\end{figure}

\begin{figure*}[htbp]
\centering
\includegraphics[width = 0.9\linewidth]{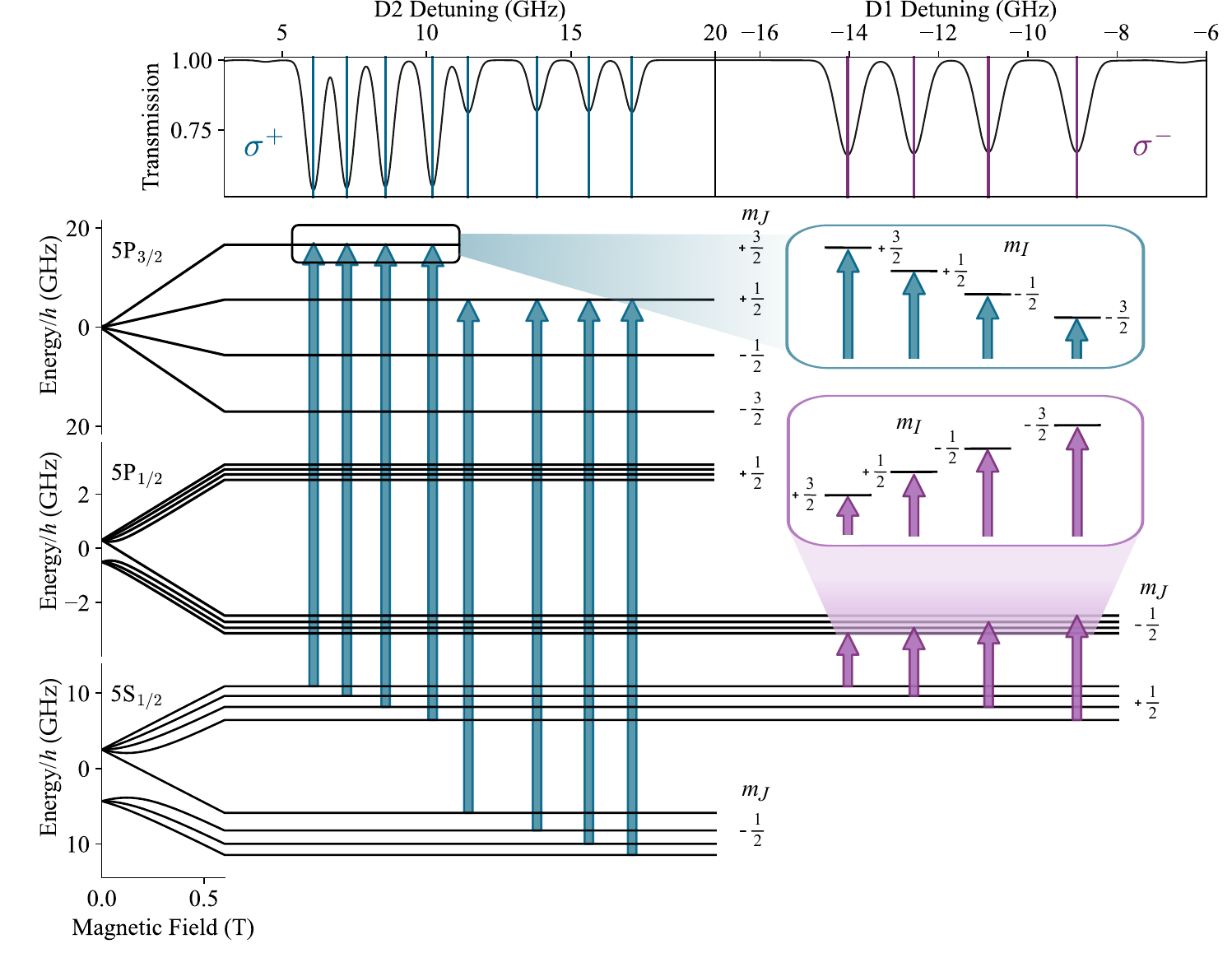}
\caption[Experimental setup for collisional transfer investigation.]{Energy levels involved in the Rb D2 (D1) transitions at 0.6 T, which are between the 5S$_{1/2}$ and 5P$_{3/2}$ (5P$_{1/2}$) energy levels. The $m_{J}$ and $m_{I}$ are labelled accordingly. The absorption spectra of linearly polarised light resonant with the D2 (D1) transitions at 75$^{\circ}$C is shown in the top panel. The light blue (purple) arrows, and corresponding lines above, mark $\sigma_{+}$ ($\sigma_{-}$) transitions excited by left-- (right--) hand circularly polarised light~\cite{f2f}.}
\label{fig:bigdiagram}
\end{figure*}

The experimental setup is shown in Fig.\,\ref{fig:Fig1}~b). We use a 2\,mm long, isotopically enriched 98$\%$ $^{87}$Rb vapour cell, which is nominally ``buffer-gas free'', in an axial magnetic field of 0.6\,T, produced by two cylindrical ``top hat'' NdFeB permanent magnets. The field is uniform across the length of the cell at the 0.1$\%$ level~\cite{WhitingEIT}. The central transmission frequency of the etalon filter is referenced using a laser resonant with the D1 (795\,nm) Rb absorption line. Left-handed circularly polarised D2 (780\,nm) light is aligned through the vapour cell for buffer-gas induced D1 fluorescence detection. A  lens of focal length 200\,mm focuses the beams to waists of \SI[separate-uncertainty]{100\pm5}{\micro\meter} $\times$ \SI[separate-uncertainty]{78\pm5}{\micro\meter} (780\,nm) and \SI[separate-uncertainty]{65\pm5}{\micro\meter} $\times$ \SI[separate-uncertainty]{90\pm5}{\micro\meter} (795\,nm) inside the cell. After the vapour cell, the output light is re-collimated using a lens of focal length 200\,mm and split into two paths using a 50:50 beam splitter cube. An interference filter is inserted into each path; in the reflected path of the beam splitter cube, the excitation (input) light is extinguished, leaving only the light produced from collisional transfer, while the opposite occurs for the transmission path. This set-up has the advantage that several signals can be monitored simultaneously. First, the transmission of the input D2 light can be monitored on a photodiode (PD1) whilst the laser frequency is scanning; this trace serves as a reference marker. Second, the light produced from collisional transfer (i.e. D1 light) is coupled down a multimode fibre, the output of which is divided into two by a 50:50 beam splitter such that the D1 fluorescence can be monitored on two separate photon counters (PC1 and PC2). The bandwidth of the interference filters are much broader than the widths of the resonance features (see Fig.~\ref{fig:bigdiagram} top panel as an example), hence the need for a second ultra narrow-band (typically narrower than the resonance features) filter for fine spectral resolution. We use an etalon filter (EF) on the optical path before PC2. An EF consists of a high-reflection coated plano-convex lens held in a temperature stabilised mount (see inset to Fig.~\ref{fig:Fig1}~b)). It has a full-width-at-half-max (FWHM) of 130~MHz and a central frequency that can be tuned by altering its temperature. The central frequency is stable to 10~MHz over a 2 hour period~\cite{HigginsEtalon}. Inputting D1 light through the set-up, before fluorescence measurements using D2 light are recorded by the photon counters, enables the central frequency of the etalon to be aligned onto one of the D1 resonance features. An example of the etalon transmission, which is detected on photodiode PD2, and the D1 resonance features, which is detected on photodiode PD1 when the interference filter is removed, is shown in Fig.~\ref{fig:Fig3}~a). Once the EF has been tuned to the correct resonance and has stabilised, the interference filter before PD1 is replaced and the D1 input light is replaced with D2 input light ready for the experiment to commence.

The vapour cell is heated to 75$^\circ$C, which is required to increase the Rb number density in the vapour cell to get appreciable absorption. When the temperature is too high, the medium becomes optically thick and we see saturation effects. To understand the collisional processes that occur inside the vapour cell, we must consider when the Rb atoms are likely to collide with a buffer gas atom (inter-species collisions) and when they might collide with another Rb atom (intra-species collisions). At low temperatures ($\textless$ 86$^\circ$C), as shown in Fig.\ref{fig:Fig1}~c), the dominating collisions are between Rb and buffer gas atoms. Inter-species collisions can transfer the Rb atom to: the other 5P state (inter-manifold); to a different level within the same 5P state (intra-manifold); or back down to the ground state (quenching). At higher temperatures (86$^\circ$C-110$^\circ$C), we also see Rb-Rb collisions~\cite{weller2011absolute}; these can cause the same type of transfer processes as stated in the inter-species case. 

In order to understand the processes occurring in the collisions we need a clear picture of the energy levels in $^{87}$Rb; this enables us to understand which transitions are available to us, and what wavelengths and polarisations of light will excite (or be produced by) these transitions. 

Fig.~\ref{fig:bigdiagram} shows the $\sigma_{+}$ D2 transitions, which has a wavelength of 780~nm, and the $\sigma_{-}$ D1 transitions, which has a wavelength of 795~nm. The effect of the 0.6~T magnet on the 5S$_{1/2}$, 5P$_{3/2}$, and 5P$_{1/2}$ state energy levels are shown. In zero-field $F$ and $m_{F}$ are good quantum numbers, however when we move into the HPB regime, the states split and regroup such that the good quantum numbers are now $m_{I}$ and $m_{J}$. These states are grouped in sets of four levels, with a common $m_{J}$, with each state having a different $m_{I}$. The $m_{I}$ and $m_{J}$ states are labelled in Fig.~\ref{fig:bigdiagram} accordingly. 
The absorption spectra of linearly polarised light, with $\overrightarrow{k} \parallel \overrightarrow{B}$, resonant with the D2 (left) and D1 (right) is shown in the top panel, respectively. The light blue (purple) arrows, and corresponding lines above, mark $\sigma_{+}$ ($\sigma_{-}$) transitions excited by left- (right-) hand circularly polarised light~\cite{f2f}.

\section{Experimental Results}
\label{sec:results}
\subsection{Spectral filtering the output fluorescence}

\begin{figure}[htpb]
\centering
\includegraphics[width = \linewidth]{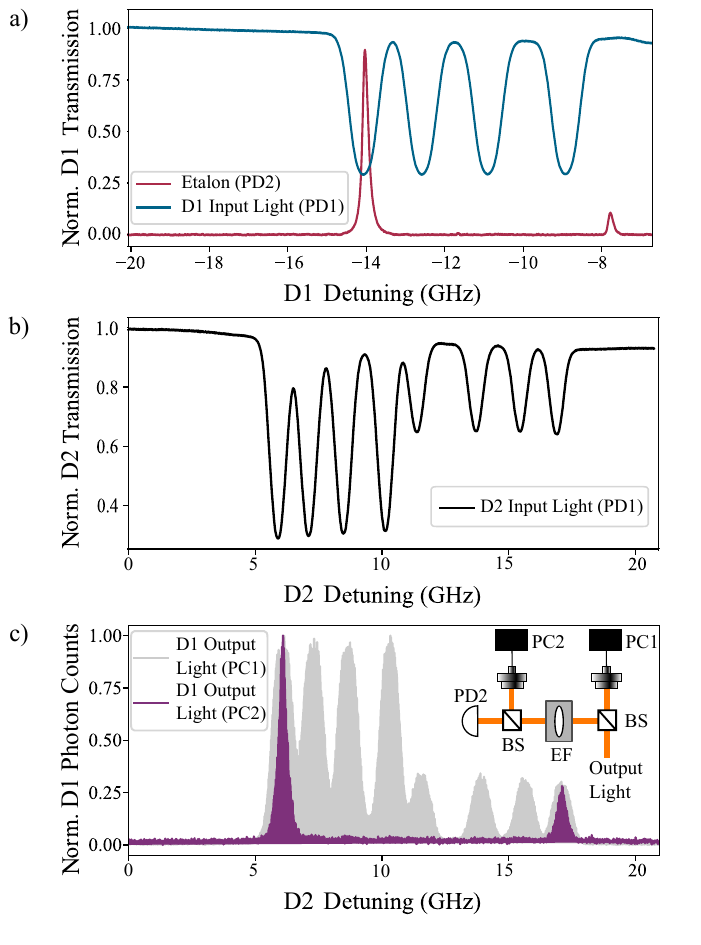}
\caption[Collisional transfer through etalon filter, scanning \SI{780}{\nm} laser, fix etalon filter on D1 resonance line.]{D2 input laser light, D1 fluorescence collected. At \SI{75}{\celsius}, in the regime where collision rate is first order with respect to Rb concentration and Rb--buffer gas collisions dominate. Panel a) shows the etalon transmission profile on a \SI{795}{\nm} laser scan relative to the D1 absorption lines. Here the etalon is positioned on the left most transition ($m_J~=~1/2, m_I~=~3/2 \rightarrow m_I~=~-1/2, m_I~=~3/2$). Panel b) shows a \SI{780}{\nm} laser scan over the D2 transition lines. Panel c) shows the fluorescence detected when the laser scan is that of panel b). Both yellow (without etalon) and purple (with etalon) traces show fluorescence that has passed through a narrow band interference filter with a central transmission frequency of \SI{795}{\nm}. Zero probe detuning for panel a) (panels b) and c)) is the weighted D1 (D2) line centre of naturally abundant Rb in zero magnetic field~\cite{Siddons_2008}.}
\label{fig:Fig3}
\end{figure}

Fig.~\ref{fig:Fig3} shows the complete dataset when we scan the frequency of the D2 input light, while the etalon filter central frequency has been positioned on one of the D1 hyperfine transmission frequencies. In this example, the etalon filter has been centred on ($m_J = 1/2, m_I = 3/2 \rightarrow m_I = -1/2, m_I=3/2$), as shown in Fig.~\ref{fig:Fig3}~a). In Fig.~\ref{fig:Fig3}~b) the transmission spectrum of the D2 input laser light is shown, while Fig.~\ref{fig:Fig3}~c) shows the normalised detection rate of the D1 photons as the D2 laser frequency is scanned. The grey trace in panel c) shows the D1 fluorescence collected on a photon counter (PC2) without the narrowband etalon filtering; this trace contains no spectral information of the D1 photons other than they are in the range 793-797 nm (due to the interference filter in the collecting path transmitting at (795$\pm$2)~nm. As expected, the rate of D1 photon production is higher when more D2 resonant laser light is absorbed by the medium. 

The inclusion of the etalon filter provides higher-resolution spectral information. From panels a) and b) we learn that when D2 light is at the frequency of the left-most D2 hyperfine transition, the D1 photons produced by the medium, via collisional transfer, are at the frequency of the left-most D1 hyperfine transition, which is a decay from an excited state 5P$_{1/2}$ with $m_J = 1/2, m_I = 3/2$. There are only two D2 input detunings which cause the production of these D1 photons: these detunings correspond to exciting the atoms into the 5P$_{3/2}$ $m_J = 3/2, m_I = 3/2$ and $m_J = 1/2, m_I = 3/2$ excited states. 

\begin{figure}[htpb]
\centering
\includegraphics[width = \linewidth]{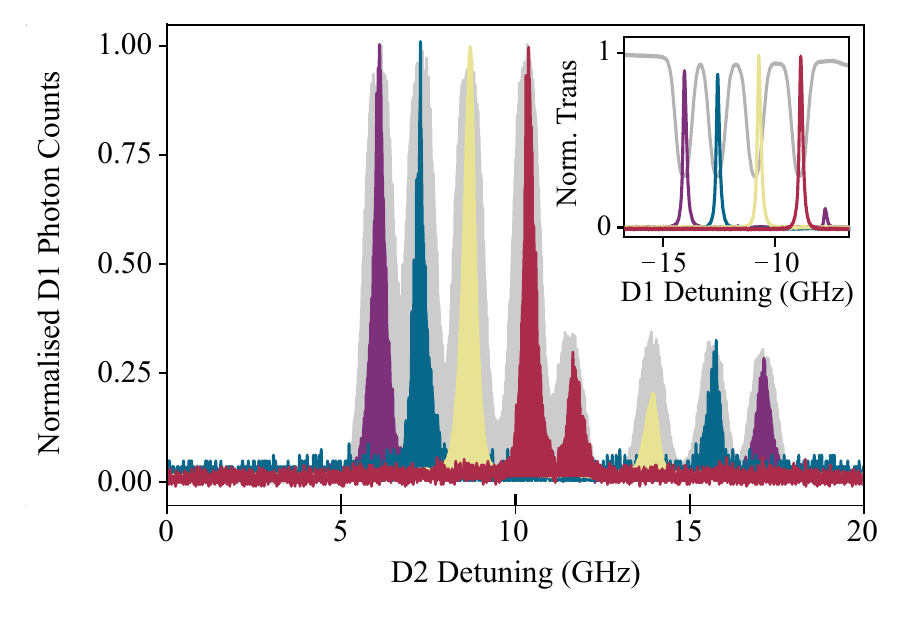}
\caption[Scan over D2 lines, fix etalon filter on different D1 absorption lines.]{D2 scanning laser light in, D1 fluorescence output light. Inset: The etalon is aligned to transmit on, from left-to-right, $|-1/2, 3/2 \rangle \rightarrow |1/2, 3/2\rangle $ transition (purple), $|-1/2, 1/2 \rangle \rightarrow |1/2, 1/2\rangle $ (blue), $|-1/2, -1/2 \rangle \rightarrow |1/2, -1/2\rangle $ transition (yellow) and  $|-1/2, -3/2 \rangle \rightarrow |1/2, -3/2\rangle $ (red). Zero probe detuning is the weighted D2 line centre of naturally abundant Rb in zero magnetic field~\cite{Siddons_2008}.}
\label{fig:Fig4}
\end{figure}

Translating the central frequency of the etalon transmission to a different D1 transition changes the fluorescence spectrum, which is evident in Fig.~\ref{fig:Fig4}. Viewing this in combination with the energy level diagram in Fig.~\ref{fig:bigdiagram}, we summarise the transitions in Table~\ref{tab:transitions}.

\begin{table}[htbp]
\centering
\caption{State change during collisions}
\def\arraystretch{2.5}
\begin{tabular}{|c|c|}
\hline
D2 excitation transition & D1 fluorescence transition \\($ |m_J, m_I\rangle \rightarrow |m_J', m_I'\rangle$) & ($ |m_J', m_I'\rangle \rightarrow |m_J, m_I\rangle$)\\
\hline
\makecell{$ |1/2, 3/2\rangle \rightarrow |3/2, 3/2 \rangle $\\ $|-1/2, 3/2\rangle \rightarrow | 1/2, 3/2\rangle $} & $|-1/2, 3/2 \rangle \rightarrow |1/2, 3/2\rangle $\\
\makecell{$ |1/2, 1/2\rangle \rightarrow |3/2, 1/2 \rangle $\\ $|-1/2, 1/2\rangle \rightarrow | 1/2, 1/2\rangle $} & $|-1/2, 1/2 \rangle \rightarrow |1/2, 1/2\rangle $\\
\makecell{$ |1/2, -1/2\rangle \rightarrow |3/2, -1/2 \rangle $\\ $|-1/2, -1/2\rangle \rightarrow | 1/2, -1/2\rangle $} & $|-1/2, -1/2 \rangle \rightarrow |1/2, -1/2\rangle $\\
\makecell{$ |1/2, -3/2\rangle \rightarrow |3/2, -3/2 \rangle $\\ $|-1/2, -3/2\rangle \rightarrow | 1/2, -3/2\rangle $} & $|-1/2, -3/2 \rangle \rightarrow |1/2, -3/2\rangle $\\
\hline
\end{tabular}
  \label{tab:transitions}
\end{table}

We deduce that during the collisional transfer process, which transfers an atom from the 5P$_{3/2}$ state to the 5P$_{1/2}$ state, the $m_{J}$ quantum number of the atom can change, but the nuclear spin projection quantum number, $m_{I}$, is preserved. 

\subsection{Spectral profile of the output fluorescence}

\begin{figure}[htpb]
\centering
\includegraphics[width = \linewidth]{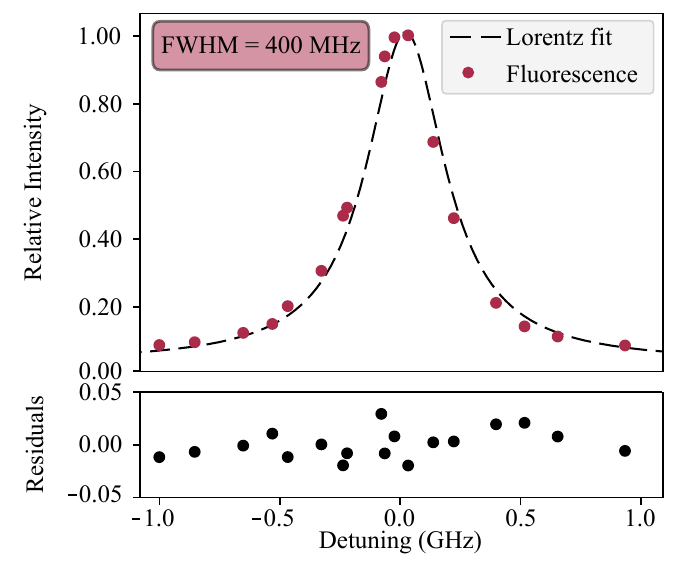}
\caption[Spectral profile of D1 fluorescence with Lorentzian fit.]{ Upper panel shows the relative intensity of D1 fluorescence (red dots) as the etalon transmission frequency is altered, with a Lorentzian fit to the measured fluorescence rate profile (black dashed line). This is the convolution of the etalon transmission profile (\SI{130}{\MHz} Lorentzian) and the fluorescence profile emitted by the atoms. Zero detuning is the central resonance frequency of the $m_I = 3/2$ $\sigma_-$ D1 absorption line. Residuals are shown beneath. }
\label{fig:moveetalon}

\end{figure}

Not only can analysis with the etalon filter inform us of which states the collisional transfer populates, it can also be used to determine the spectral characteristics of the emitted fluorescence, namely the lineshape and the linewidth. Fig.~\ref{fig:moveetalon} demonstrates the spectral distribution of the produced D1 fluorescence by fixing the D2 laser frequency and moving the central frequency of the etalon transmission in the observation path. The measured profile is a good fit to a  Lorentzian lineshape, and because each fluorescence-measurement point measures a range of frequencies given by the profile of the etalon, the measured profile is a convolution of the etalon profile and the profile of the emitted fluorescence. The etalon profile is known to be a Lorentzian~\cite{HigginsEtalon}, and it can be shown that the convolution of two Lorentzians produces a third Lorentzian with a FWHM of $\Gamma_{\mathrm{f}}$~=~$\Gamma_{\mathrm{1}}+\Gamma_{\mathrm{2}}$~\cite{loudon2000quantum}. Therefore we determine that the  fluorescence spectrum is also close to Lorentzian, and that its width is the difference between the measured width and the width of the etalon filter used to take the measurement. Therefore we conclude that the fluorescence is approximately Lorentzian with a FWHM of approximately 270~MHz.

\section{A collision model and comparison with experiment}
\label{sec:model}
\subsection{A simple kinematic model}
To better understand the fluorescence spectral profiles we observe, we create a basic model of the collisions in the medium. We use a simple hard-sphere-collisions picture~\cite{youngandfreedman}, which we expect to be valid as we are in the classical scattering regime. We use a simple Monte-Carlo model to simulate the collisions, and to explain the FWHM and lineshape of the emitted fluorescence (as shown in Fig.~\ref{fig:moveetalon}).  We carry out these operations array-wise, and model 10 million collisions, which runs in approx 30 seconds on an Intel Core i5 processor. The model results shown in this section use a buffer gas of molecular nitrogen, N$_2$. Other likely candidates for residual buffer gas in a cell, such as CH$_4$, He, Ne, Ar, are all considerably lighter than Rb, so choosing a different buffer gas has a negligible effect on the final results.

For each collision we initialise the $x$, $y$ and $z$ velocities of both (Rb and buffer gas) atoms. In the experiment (measuring the spectral profile of the fluorescence) we excite with a resonant D2 (\SI{780}{\nm}) beam directed along the $z$-axis, so only interact with atoms which have $v_z < \Gamma/k_z$~\cite{hughes2018velocity}, where $\Gamma$ is the natural linewidth of the excitation transition, and $k_z$ the $z$-component of the light wavevector (though in this case $k = k_z$). In the model, $v_z$ of all Rb atoms is set to be \SI{0}{\m/s}, and all other velocity components are randomly chosen from a Gaussian distribution at $T = $\,\SI{75}{\celsius}. We use a simple spheres colliding picture (like that illustrated in Fig.~\ref{fig:Fig1}) to calculate the velocities of both particles after the collision. To do this we also randomly select an impact angle for each collision, and from this calculate the contact normal vector, $\Vec{n}$. We then calculate the relative velocity along the contact vector
\begin{equation}
    v_{\text{rel}} = (\Vec{v}_\text{Rb} - \Vec{v}_\text{Buff}) \cdot \Vec{n}.
\end{equation}
The velocities after the collision, with i denoting the `intermediate' state of our calculation, are given by 
\begin{equation}
    \Vec{v}_\text{Rb,i} = \Vec{v}_\text{Rb} -  v_\text{rel}  \frac{2 m_\text{Buff}}{m_\text{Buff}+m_\text{Rb}} \Vec{n},
\end{equation}
and
\begin{equation}
    \Vec{v}_\text{Buff,i} = \Vec{v}_\text{Buff} +  v_\text{rel}  \frac{2 m_\text{Rb}}{m_\text{Buff}+m_\text{Rb}} \Vec{n}.
\end{equation}

These solutions are the velocities of both particles after a perfectly elastic collision, and do not take into account the energy change from the state change which occurs during the collision. 

During the collision the internal energy of the Rb atom decreases as it transitions to a lower energy state. We propose that this energy is transferred to the kinetic energy of the colliding particles. A collision in which energy in released, and the final total kinetic energy of the particles exceeds the initial kinetic energy, is known as a `superelastic' collision~\cite{herrero1972superelastic} and has a coefficient of restitution, $e>1$~\cite{duan2023theoretical}. We use this concept to include the state change into the kinetic model as follows. The calculation is carried out in the centre-of-mass (cm) frame
\begin{equation}
    \Vec{v}_\text{cm} = \frac{m_\text{Rb}\Vec{v}_\text{Rb,i} + m_\text{Buff}\Vec{v}_\text{Buff,i}}{m_\text{Rb}+m_\text{Buff}},
\end{equation}
and final velocities are given by 
\begin{equation}
    \Vec{v}_\text{Rb,f} = (\Vec{v}_\text{Rb,i} - \Vec{v}_\text{cm}) \cdot e + \Vec{v}_\text{cm},
\end{equation}
and 
\begin{equation}
    \Vec{v}_\text{Buff,f} = (\Vec{v}_\text{Buff,i} - \Vec{v}_\text{cm}) \cdot e + \Vec{v}_\text{cm}.
\end{equation}
Again, i denotes the previously calculated intermediate velocities, f the final velocities, and $e$ is the coefficient of restitution.

The extra energy we are adding in the collision is $\Delta E$, the energy difference between the 5P$_\text{3/2}$ and 5P$_\text{1/2}$ states. At our operating temperature, \SI{75}{\celsius}, this energy is very similar to the thermal energy of the atoms, $k_\text{B}T \approx \Delta E$. The coefficient of restitution, $e$, is numerically calculated to conserve the total energy over a large number of collisions, $n$, such that

\begin{multline}
   \sum_n \left[\frac{1}{2}m_{\text{Rb}}v_{\text{Rb,i}}^2 + \frac{1}{2}m_{\text{Buff}}v_{\text{Buff,i}}^2 + \Delta E \right]=  \\ \sum_n \left[\frac{1}{2}m_{\text{Rb}}v_{\text{Rb,f}}^2 + \frac{1}{2}m_{\text{Buff}}v_{\text{Buff,f}}^2\right].
\end{multline}

This results in a value of $e = 1.3$. 
\begin{figure}[htpb]
\centering
\includegraphics[width = 1\linewidth]{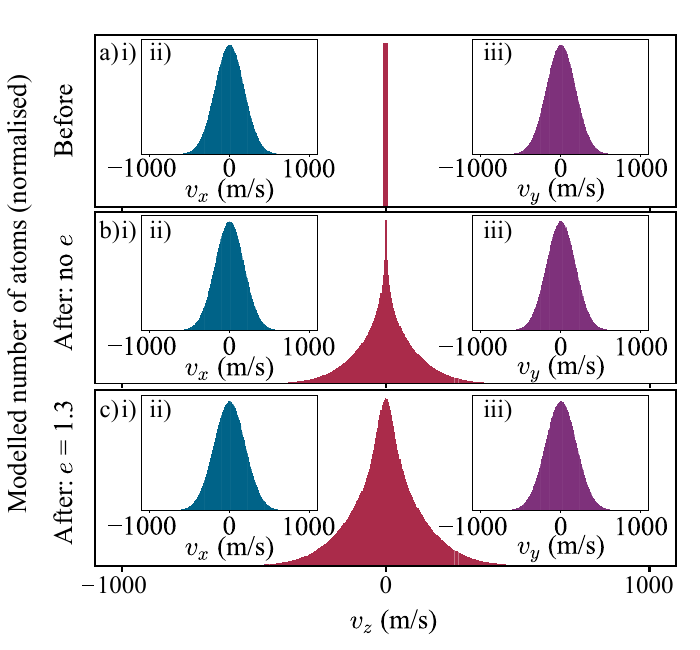}
\caption[Velocity histograms before and after collision.]{Simulated D1 (\SI{795}{\nm}) atomic velocity distributions along i) $z$, ii) $x$ and iii) $y$ before the collision (a), and after the collision, without (b) and with (c) the extra energy from the state change via a coefficient of restitution of 1.3.}
\label{fig:collisions:velocity-distributions}
\end{figure}

We obtain a fluorescence lineshape from the calculated velocities by histogramming the final Rb velocities along the observation axis (initially $v_z$), and converting velocity to detuning, via $\Delta \nu = v_z / \lambda$. The fluorescence lineshape in this direction is independent of$v_x$ and $v_y$. Fig.~\ref{fig:collisions:velocity-distributions} shows the modelled velocity distributions along i) $z$, ii) $x$, and iii) $y$. Distributions are plotted before  (a), and after the collision, without (b) and with (c) the extra energy from the state change via a coefficient of restitution of 1.3. It can be seen that the distributions in $x$ and $y$, which start Gaussian, are changed very little by the collision, though they are broadened slightly when the extra energy is included: c)\,ii) and c)\,iii). The distribution in $z$, on the other hand, changes significantly. Initially all atoms have $v_z$\,=\,0, as shown in  Fig.~\ref{fig:collisions:velocity-distributions}\,a)\,i). After the collision, when the extra energy is not included as in Fig.~\ref{fig:collisions:velocity-distributions}\,b)\,i), the distribution still has a high narrow peak at $v\,=\,0$, and a prominent  cusp. Including the extra energy via the coefficient of restitution $e=1.3$ gives the distribution in Fig.~\ref{fig:collisions:velocity-distributions}\,c)\,i), which is broader, less cusped, and closer to Lorentzian. This lineshape is in agreement with previous studies where velocity changing collisions have been observed to have cusped lineshapes~\cite{PhysRevA.43.1366,doi:10.1143/JPSJ.72.1936} and a cusped lineshape collision kernel has been described~\cite{PhysRevLett.108.183202}.

\subsection{Comparison with experiment}

To compare the theoretical model with the lineshape we measure in the experiment, the velocity distribution in Fig.~\ref{fig:collisions:velocity-distributions}\,c)\,i) is converted to a frequency profile and convolved with the known etalon filter profile. 

\begin{figure}[htpb]
\centering
\includegraphics[width = \linewidth]{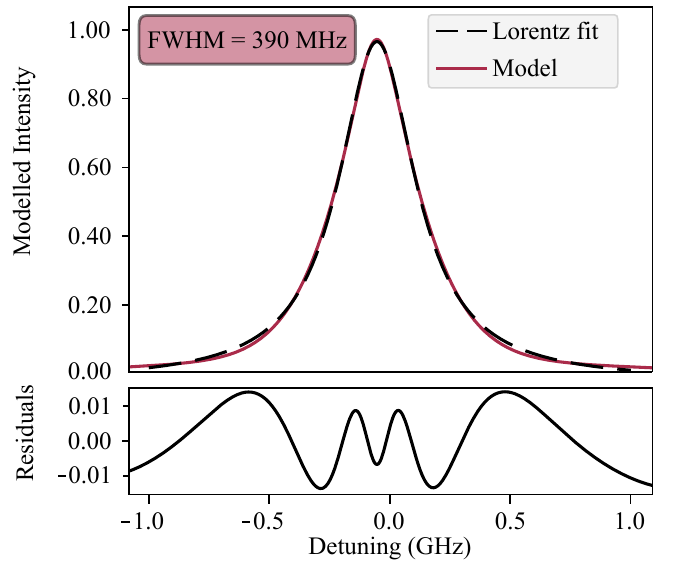}
\caption[Model fit to Lorentzian after convolution with filter]{Fit of the model to a Lorentzian, after convolution with the \SI{130}{\MHz} Lorentzian etalon filter profile. The width of this profile is \SI{390}{\MHz}, and is a very good fit to the Lorentzian.}
\label{fig:model-resonance}
\end{figure}

Fig.~\ref{fig:model-resonance} shows a fit of the model to a Lorentzian, after convolution with the \SI{130}{\MHz} Lorentzian filter profile. The width of this profile is now \SI{390}{\MHz}, which matches very closely the \SI{400}{\MHz} measured profile as shown in Fig.~\ref{fig:moveetalon}. The fit is excellent, with residuals of $<$1\%.

\section{Fine-structure changing collisions with different buffer gases and alkali metals.}
\label{sec:different}
\subsection{Different buffer gases with Rb}
The vapour cell used in the studies discussed so far was nominally ``buffer-gas free''. However, we have shown that there is some additional gas present due to collisional transfer. Since the additional gas, or gases, in our vapour cell is unknown, we have repeated the studies discussed previously with vapour cells with known buffer gases and concentrations. These cells are cubes with side lengths of \SI{1}{\mm} and contain methane and molecular hydrogen, which is known to produce a combined additional broadening of 24 MHz~\cite{weller2013absolute}. One of these cells also contains a large amount of helium, which provides another 300 MHz additional broadening~\cite{keaveney2019quantitative, ponciano-ojeda2021stokes}. 

\begin{figure}[htpb]
\centering

\includegraphics[width = 0.95\linewidth]{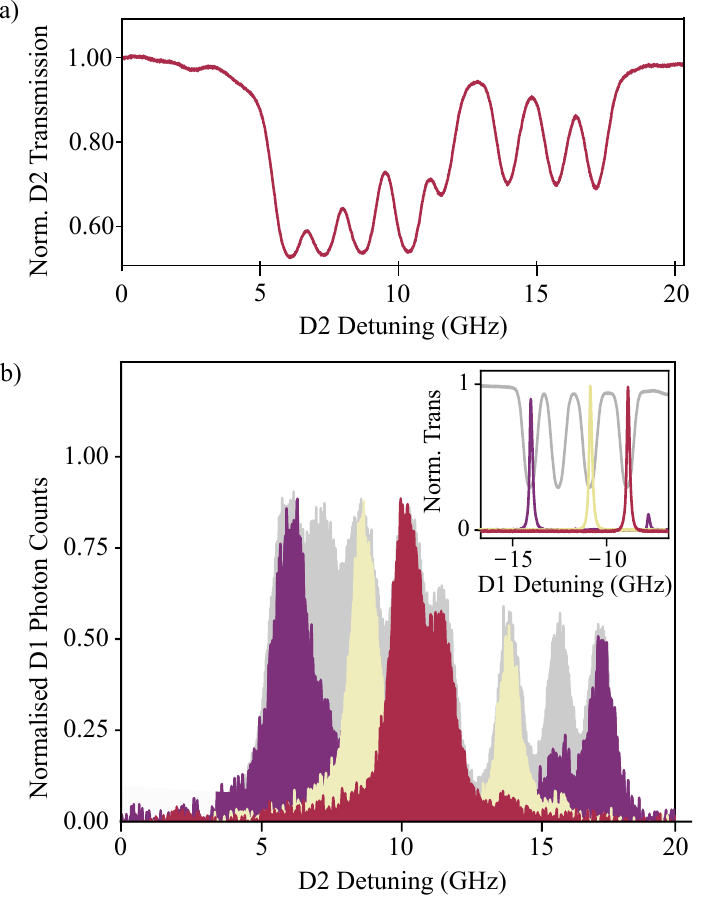}
\caption[Collisional transfer in helium broadened cell. ]{In \SI{300}{\MHz} He-broadened cell. D2 scanning laser light in, D1 fluorescence output light. a) D2 transmission spectrum. b) Fluorescence without etalon (grey) and with etalon (purple, yellow, red) transmitting on $|-1/2, 3/2 \rangle \rightarrow |1/2, 3/2\rangle $ transition, $|-1/2, -1/2 \rangle \rightarrow |1/2, -1/2\rangle $ transition and  $|-1/2, -3/2 \rangle \rightarrow |1/2, -3/2\rangle $, respectively, as shown in the inset. Zero probe detuning is the weighted D2 line centre of naturally abundant Rb in zero magnetic field.}
\label{fig:He-cell}
\end{figure}

The fluorescence spectra for the He-broadened cell are shown in Fig~\ref{fig:He-cell}.  The D2 input light is scanned over the $\sigma_+$ absorption lines (transmission spectrum shown in Fig~\ref{fig:He-cell}~a)) and the D1 fluorescence is shown in panel b) with and without the etalon filter. Shown inset is the position of the central frequency of the etalon transmission with respect to the D1 absorption features. 

As previously, spectra  were taken in the low temperature regime, at \SI{75}{\celsius}. We observed the same $m_I$ conservation pattern in both cells, despite the additional \SI{300}{\MHz} broadening meaning the transition peaks are no longer well distinguished. We would expect the temperature--collision rate graph, equivalent to Fig.~\ref{fig:Fig1}~c) for the main vapour cell, to switch from the linear to the quadratic regime at a different temperature, because the buffer gas number density is higher, and the different buffer gas will have a different collisional cross-section. Therefore the crossover point, where Rb--Rb state changing collisions happen at a higher rate than Rb--buffer state changing gas collisions will occur at a higher different temperature. 

\subsection{Different alkali metals}
Rb happens to have a fine structure splitting approximately equal to $k_\text{B}T$ at \SI{75}{\celsius}. Table~\ref{tab:energy-gaps} shows how the splitting varies in alkali metal atoms at \SI{100}{\celsius}. Heavy alkali atoms have a smaller ratio meaning the collisional transfer is less favourable. In the lighter elements, such as K and Na however, the ratio increases. In experiments where the presence of fine-structure changing collisions can have deleterious effects, such as heralded-single-photon generation, Cs would be less affected than the lighter metals. By contrast, fine structure changing collisions are likely to be prominent in vapour cells with the lighter metals,  and the difference in wavelength of the D2 and D1 transitions is also smaller making it more difficult to separate with standard interference filters. This is particularly relevant for solar physics applications, where heated vapour cells of Na and K are used in magneto-optical filters~\cite{refId0, refId1, refId2}. 
  
\begin{table}[tbh]
    \centering
    \begin{tabular}{| c| c |c |}
         \hline
        Atom & $k_\text{B} T/\Delta E$ at \SI{100}{\celsius}\\
         \hline
         Na & 15\\
        K & 4.5\\
        Rb & 1.1\\
        Cs & 0.48\\
 
        \hline
\end{tabular}
\caption{Table of ratios of thermal energy to P-state fine-structure splitting in alkali metal atoms, at \SI{100}{\celsius}.}
\label{tab:energy-gaps}
\end{table}

\section{Conclusions and Outlook}
\label{sec:conc}

Using an etalon filter we were able to obtain high-resolution spectral information about fine-structure changing collisions in $^{87}$Rb upon D2 excitation in the hyperfine Paschen-Back regime. Our data show that  during the collisional transfer process 5P$_{3/2}\rightarrow 5{\rm P}_{1/2}$, the $m_{J}$ quantum number of the atom  changes, but the nuclear spin projection quantum number, $m_{I}$, is conserved, as expected. A simple kinematic model incorporating a coefficient of restitution in the collision accounted for the change in velocity distribution of atoms undergoing collisions, and the resulting fluorescence lineshape.
   
The vapour cell used in this investigation was nominally ``buffer-gas free''.  We have shown that using photon-counting apparatus to detect fine-structure collisions provides a sensitive method to detect the presence of buffer-gas in such cells. When there is a large amount of buffer-gas present the additional pressure broadening can be measured from a fit to a Voigt
profile~\cite{Zentile2015b}. Measuring fine-structure changing collisions by monitoring fluorescence could be used to measure the buffer-gas pressure, in the low-buffer-gas regime  where the additional line broadening would be difficult to measure.



\noindent\textbf{Data Availability.}
Data underlying the results presented in this paper are available from \cite{dataavailability}. The data that supports the findings of this study are openly available at the following URL/DOI: https://doi:10.15128/r1rr171x31q.
~\\
~\\
\noindent\textbf{Acknowledgements.} The authors thank Renju S. Mathew for preliminary work on this investigation,  Jeremy M. Hutson and Lina M. Hoyos-Campo for fruitful theoretical discussions, Paul F. Griffin for discussions regarding the potential of using this technique to measure buffer-gas pressures in vapour cells, and Liam A. P. Gallagher for constructive comments on the manuscript.
~\\
~\\
\noindent\textbf{Funding.} The authors thank EPSRC for funding this work (Grant No. EP/R002061/1). For the purpose of open access, the authors have applied a Creative Commons Attribution (CC BY) licence to any Author Accepted Manuscript version arising from this submission.
~\\
~\\
\noindent\textbf{Conflict of interest} The authors declare no conflicts of interest.

\bibliographystyle{unsrt}
\bibliography{references}


\end{document}